\documentclass[%
 aip,
 amsmath,amssymb,
 reprint,%
]{revtex4-1}

\usepackage{graphicx}
\usepackage{dcolumn}
\usepackage{bm}

\usepackage[utf8]{inputenc}
\usepackage[T1]{fontenc}
\usepackage{mathptmx}
\DeclareMathOperator*{\argmax}{arg\,max}
\DeclareMathOperator*{\argmin}{arg\,min}
\begin{document}

\preprint{AIP/123-QED}

\title[Entropic Regression for Neurological Applications]{Entropic Regression for Neurological Applications}

\author{Jeremie Fish}

 
 \email{fishja@clarkson.edu}
\author{Alexander DeWitt}%
 

\author{Abd AlRahman R. Almomani}
\affiliation{Department of Electrical and Computer Engineering, Clarkson University.}
  \affiliation{Clarkson Center for Complex Systems Science ($C^3S^2$).}
\author{Paul J. Laurienti}
\affiliation{Department of Radiology, Wake Forest School of Medicine}
\author{Erik Bollt}
\affiliation{Department of Electrical and Computer Engineering, Clarkson University.}
  \affiliation{Clarkson Center for Complex Systems Science ($C^3S^2$).}

\date{\today}

\begin{abstract}
 The ultimate goal of cognitive neuroscience is to understand the mechanistic neural processes
underlying the functional organization of the brain. Key to this study is understanding structure of both the structural and functional connectivity between anatomical regions. In this paper we follow previous work in developing a simple dynamical model of the brain by simulating its various regions as Kuramoto oscillators whose coupling structure is described by a complex network. However in our simulations rather than generating synthetic networks, we simulate our synthetic model but coupled by a real network of the anatomical brain regions which has been reconstructed from diffusion tensor imaging (DTI) data. By using an information theoretic approach that defines direct information flow in terms of causation entropy (CSE), we show that we can more accurately recover the true structural network than either of the popular correlation or LASSO regression techniques. We demonstrate the effectiveness of our method when applied to data simulated on the realistic DTI network, as well as on randomly generated small-world and Erd{\"o}s-R{\'e}nyi (ER) networks.
\end{abstract}

\maketitle

\begin{quotation}
The field of cognitive neuroscience seeks to understand the function of the brain, as related to the physical brain structure. Knowledge of the connectivity between functional regions is central to this comprehension. However, these relationships are unknown a priori. Well known methods have been adopted for this deduction from time series, including correlation and LASSO regression. Such methods, especially LASSO, can lead to recovery of a majority of true connections with sufficient sample size. However as we  show, LASSO also infers a large fraction of false connections which do not exist in the true network. To circumvent this issue we utilize our recenty developed entropic regression.  Entropic regression is an information theoretic technique that is especially well suited for this problem, as it allows for  optimal selection of basis functions as related to the underlying information flow of the dynamical system. We show that entropic regression yields high recovery of true edges, while simultaneously limiting the number of falsely inferred connections, thus associated with excellent ROC performance (receiver operating characteristic curve). 

\end{quotation}

\section{Introduction}

Complex networks are all around us, from structural networks such as roads and flights \cite{zhan1998,morrell2007}, to social networks such as facebook and twitter \cite{lewis2008,suh2010} to chemical networks, such as Belousov-Zhabotinsky oscillators \cite{tompkins2015, torbensen2017}, biological systems \cite{desilva, mason} and network neuroscience \cite{sporns, bassett2006,bassett2011} which we will be studying extensively here. We will consider a description of the brain as partitioned into 83 anatomical regions \cite{bonilha2015}, whose function of these interacting elements is to be described as a complex network structure.
\paragraph*{}
Knowledge of the structure of the complex networks helps to determine how they will respond to stresses, such as dynamical perturbations \cite{barzel2013, wang2016}, or to structural network perturbations \cite{pas1997,korilis1999}. However frequently the underlying network structure of a system is unknown. What may be available to us is time series data collected at each node.  In the case of functional magnetic imaging (fMRI) data, time varying intensity is associated with each of the voxels, as a three-dimensional movie.  The associated metabolism of an active region of the brain becomes apparent by increased oxygen levels resulting the the blood oxygenation level dependent (BOLD) signal that can be inferred by magnetic resonance \cite{ogawa1993}. These perturbations and interacting variations carry information about the underlying network  connectivity of the brain.  However, the inverse problem of inferring  the underlying network connectivity from observations of state, is inherently an ill-posed problem that is sometimes called network tomography, and it remains a difficult problem today.
\paragraph*{}
In this work we utilize a recently developed method named entropic regression \cite{almomani2020} to recover the underlying network structure of synthetic data which is generated from a network Kuramoto model from three different types of networks from synthetic models to a real structural brain  network: an Erd\H{o}s-R{\'e}nyi \cite{erdos1960} (ER) network as a classical random graph model, a small-world network following \cite{stoltz2017} with designed structure, and the experimentally observed DTI network by methods as discussed below. On these network models of successive challenge and realism, we will simulate a popular synthetic dynamic to challenge the accuracy entropic regression.  We simulate  the  first order network Kuramoto model, that has popularly  been used including for the generation of synthetic fMRI data \cite{stoltz2017}. With the simulated data we demonstrate that entropic regression accurately outperforms for network recovery as compared to  other leading methods, including correlation and LASSO regression utilizing the Bayesian information criterion (BIC). 
 
\section{Materials and Methods}
 \subsection{Complex Networks}
\paragraph*{}
For this work we will use the terms graph and complex network interchangeably. A graph $\mathcal{G}=(\mathcal{V},\mathcal{E})$ is a set of nodes $\mathcal{V}$ and edges $\mathcal{E} \subseteq \mathcal{V} \times \mathcal{V}$. In context here where we will associate each anatomical region of the brain as a node, the goal is to infer interactions as information flow between these regions. The unweighted structure of a graph of $n$ nodes can be encoded by an $n\times n$  adjacency  as follows:
\begin{equation}
[A]_{ij} = 
\begin{cases}
1 \ \mbox{if} \ (i,j) \in \mathcal{E} \ \forall (i\neq j) , \\
0 \ \mbox{otherwise}.
\end{cases}
\end{equation}
A {\it weighted} directed graph $\mathcal{G}=(\mathcal{V},\mathcal{E},\mathcal{W})$ has a weight for each edge, and it can be encoded by a weight matrix, $[W]_{ij}=w_{ij}sign([A]_{ij})$, with weights $w_{ij}$.
A simple adjacency matrix is defined by a graph where no self edges occur. A {\it complete} network is defined by the presence of all possible edges.   We will generally allow for weighted directed graphs, which are well defined by a (possibly asymmetric) weight  matrix $W$, which will be our goal to find from node-level observations only.

The degree distribution  is often used to classify complex networks, even though it does not uniquely define its generative process.  Recall that the {\it degree} of a node $d_i$ is defined by $d_i = \sum \limits_{j=1}^n [A]_{ij},$ and the degree distribution is the discrete probability function of $d_i$ sampled across $i$.  The Erd{\"o}s-R{\'e}nyi \cite{erdos1960} (ER) graph, for example, is the classical random graph wherein edges are assigned at random with probability $p$. It has been shown \cite{newman2003} in the ER graph the degree distribution is Poisson in the limit of a large number of nodes. It became clear over time that real world networks tend to be highly clustered and yet have a small shortest path distance between any pair of nodes \cite{watts1998}  and this led to the development of the Watts-Strogatz (WS) "small world" graph. ER graphs do not exhibit the clustering or short path length that the small world graphs do. 
\subsection{Structural Brain Networks}
Simulated time series data were generated using an actual structural brain network but a synthetic model, thus we call the data semi-synthetic, in addition to the synthetic complex networks with synthetic model dynamics, which we call fully synthetic data. The true brain networks were generated experimentally in previously published work \cite{bonilha2015} using diffusion tensor imaging (DTI) data. The networks generated in this prior work are publicly available and "Subject1" through "Subject20" were used in the current work. Details of the image processing and network generation can be found in that manuscript but are briefly described below. The Diffusion Toolkit (FDT) in the FSL software package \cite{behrens2003} was utilized for image preprocessing and for diffusion tensor estimation. The cerebral gray matter was parcellated into 83 regions of interest (ROIs) based on the Lausanne anatomical atlas using Freesurfer software \cite{fischl2002}. There were 41 ROIs in each hemisphere and a single brainstem region. The regions were warped to the native space of each study participant based on the anatomical image and then transformed to the DTI space using FSL. Deterministic tractography was performed using the Diffusion Toolkit software. All white matter voxels were seeded and resulting fibers were assessed to determine if they connected two of the ROIs. The ROIs served as the nodes in the adjacency matrix. An edge was present between two nodes if there was at least one white matter fiber connecting the nodes. A weighted adjacency matrix was generated by counting the number of fibers connecting two nodes and normalizing this value using the length of the fibers and the area of the ROIs as previously proposed \cite{hagmann2008}. The absence of a fiber connecting two ROIs was indicated in the adjacency matrix with a  zero ($0$). 
\subsection{Regression, regularized regression, toward entropic regression}

Suppose a parametric model, $y=x \beta$ is assumed for scalar real variable $y\in {\mathbb R}$ and $x\in {\mathbb R}^n$.  Then $q$ samples, may be stated by
\begin{equation}
Y = X\beta + \epsilon,
\end{equation}
with data matrices $X\in \mathbb{R}^{q \times n}$, $Y\in {\mathbb R}^q$ and here noise is  assumed  to be normal $\epsilon \sim \mathcal{N}(0,\sigma)$, gives 
 $\beta \in \mathbb{R}^{n }$ unknown  regressors. We assume enough samples so that the problem is over-determined, $q>n$.  The goal in ordinary least squares (OLS) is to find the closest parameteric fit to the data in the sense of square norm residual, so the  minimizer of the loss function, \cite{raftery1997, golub2013}:
\begin{equation}
\min \limits_{\beta \in \mathbb{R}^{n}} || X\beta - Y||_2. \label{eq:OLS2}
\end{equation}
  A flaw of OLS however, as noted in \cite{tibshirani1996}, is that while  the OLS estimator has a low  bias, it typically has a large variance \cite{tibshirani1996}.  It tends to suffer from over-fitting if model complexity is not well chosen.
\paragraph*{}
Tikhonov regularization is a strategy to mitigate overfitting of  OLS. The regularization term introduces a penalty term to Eq. \ref{eq:OLS2}. Let:
\begin{equation}
\min \limits_{\beta \in \mathbb{R}^{n}} || X\beta - Y ||^2_2 + R(\beta), \label{eq:OLSImproved}
\end{equation}
where $R(\beta)$ is a regularization term on $\beta$. OLS paired with common types of regularization are often referred to as ridge regression and Least Absolute Shrinkage and Selection Operator (LASSO). In ridge regression, we choose,
\begin{equation}
R(\beta) =  \lambda ||\beta||_2, \label{eq:ridge}
\end{equation}
emphasizing reduced variance.  In 
 LASSO regression, we have,
\begin{equation}
R(\beta) = \lambda ||\beta||_1, \label{eq:LASSO}
\end{equation}
which emphasizes sparsity.
Tikhonov regularization in general defines a selection of a favored solution amongst infinitely many of an ill-posed problem as the philosophy of inverse problems, and in the ridge regression scenario, this sense can be thought of as filtering  noise that is related to the small but nonzero singular values of the data matrix \cite{golub1999}. 
\paragraph*{}
The choice of the value of regularity parameter $\lambda$ is an important practical aspect of regularization, as a poor choice for this value can lead to poor estimation results. There exist several methods for its choice, but here we choose the Bayesian Information Criterion (BIC) \cite{schwarz1978}, which is one of the popular choices for selecting the value of $\lambda$.  The BIC solution is considered to be similar to maximum likelihood method for asserting the dimension of a model. The BIC estimation relies on the optimizing \cite{wit2012},
\begin{equation}\label{bic1}
\mbox{BIC}(\beta,\lambda|X) = \kappa \mbox{ln}(q) - 2 \mbox{ln}(\mathcal{L}(\beta,\lambda|X))
\end{equation}
where $\kappa$ is the number of nonzero model parameters and $\mathcal{L}$ is the maximum likelihood solution of the given model with respect to the parameters $\beta$ for a given fixed value  of the parameter $\lambda$ and given the data $X$. Then the optimal value ($\lambda_{opt}$) can be found by minimizing the BIC with respect to $\lambda$ for each $\beta$, that is:
\begin{equation}\label{bic2}
\lambda_{opt} = \min \limits_{(\beta,\lambda)} \mbox{BIC}(\beta,\lambda)
\end{equation}
This formulation introduces a penalty term on making the dimension of the model too large as well as a penalty for having a model with too many parameters, which generally provides popularly pleasing estimates of the parameter $\lambda_{opt}$ when using LASSO, \cite{tibshirani1996,schwarz1978,wit2012}.
\subsection{Entropic Regression} 

Here we review entropic regression as developed in \cite{almomani2020} specializing the network discovery algorithm of optimal causation entropy \cite{sun2014,sun2015}, which is an efficient and accurate method to infer a parameterically defined model, as a system identification problem, but based on an information theoretic criterion.
In entropic regression \cite{almomani2020}, we use the conditional mutual information of time delayed measurements of the time series as to how they interact with basis functions, for an information-theoretic criterion to iteratively select relevant basis functions.  This approach is based on our prior work in causation entropy \cite{sun2014,sun2015} which is a information flow estimator that is capable of distinguishing direct versus indirect influence and in this way it is a generalization of the concept of transfer entropy \cite{schreiber2000} that is meant for just to component dynamics.   The system identification problem can be stated in matrix form:
\begin{equation}\label{eqn1}
    \dot{Z} = \Phi(Z) \Gamma 
\end{equation}
where $Z\in\mathbb{R}^{q\times n}$ as before and is the measured state variables of the $n$-dimensional system with $q$ observations, $\dot{Z}$ is the vector field estimated from $Z$, $\Phi:\mathbb{R}^{q\times n} \mapsto \mathbb{R}^{t\times K}$ is a function that maps the state variables $Z$, to the expanded set of candidate functions (not necessarily linear), and $\Gamma\in\mathbb{R}^{K\times n}$ is the parameters matrix. 
Given a basis set of functions $\Phi = \{\phi_i(Z)\}_{i=1}^{K}$, where a row vector $\Phi_j$ is given by $\Phi_j =\{\phi_1(Z_j),\phi_2(Z_j),...,\phi_K(Z_j)\}$, and $\phi_i(Z_j):{\mathbb R}^{n} \mapsto {\mathbb R}$, is a candidate function on the $n$-dimensional observation $Z_j$, that has a high flexibility of choice.

For example in this form, we can write Lorenz's equations, with its three dimensional state vector $Z=(z_1,z_2,z_3)$, as $\dot{z}_1=\sigma (z_2 - z_1), \dot{z}_2=z_1(\rho-z_3)-z_2, \dot{z}_3=z_1 z_2 - b z_1$.
Even though Lorenz is a nonlinear equation, it is a linear combination of nonlinear functions, and the linear equation implied by Eq.~(\ref{eqn1}), using the second-order power polynomial functions, $\Phi(Z=(z_1,z_2,z_3))=\{1,z_1,z_2,z_3,z_1^2,z_1z_2,z_1z_3,z_2^2, z_2z_3,z_3^2\}$, we see that we have 10 candidate functions that contain linear and nonlinear terms (functions). However, the underlying dynamic of the Lorenz system is a linear combination of these candidate functions, and we see, for example, that $\dot{z}_1$ in the Lorenz system can be found as
$\dot{z}_1=\Phi(Z)[0, ~-\sigma, ~\sigma, ~0, ~0, ~0, ~0, ~0, ~0, ~0]^{T}$. The same applies for $\dot{z}_2$ and $\dot{z}_3$.
This kind of linearization is generally known as Carleman linearization \cite{carleman1932application}. 
The reconstructed vector field using the least squares solution may be written as:
\begin{eqnarray}\label{eq:LSprojection}
V(\dot{Z},\Phi) & = &  \Phi\Phi^{\dagger}\dot{Z} \notag \\
                         & = &  \Phi L(\dot{Z},\Phi)
\end{eqnarray}
where $\Phi^{\dagger}$ is the pseudoinverse of the matrix $\Phi$.
The problem of regression for differential equation models $\Gamma$ has  recently become especially relevant for data-driven science \cite{crutchfield1987equations, yao2007modeling, wang2016data} and recently popularized by  sparse methods such as \cite{brunton2016discovering} optimization  by LASSO regression, and by information theoretic methods \cite{sun2014,sun2015,almomani2020}.
For simplicity, we will write $\Phi(Z)$ as $\Phi$ in the following discussion.
Our entropic regression has proved competitive improvements \cite{almomani2020} based on information criterion.
\paragraph{Entropic Regression Algorithm}
The entropic regression involves an optimization method to associate data to a most informative and sparse set of basis functions.  Most informative is interpreted in the sense of mutual information between basis function (observable set) and the measured data.
The underlying optimization proceeds in two stages: forward greedy search (selection) and (backward) possible elimination of basis functions, and these  are based on the conditional mutual information amongst competing observations through the selected basis functions. In the forward stage, our objective is to select the subset $s\subset \mathcal{S}=\{1,2,...,K\}$, which represent strong candidate functions. Starting from empty set $s_0 =\emptyset$, the forward selection stage can be written as:
\begin{eqnarray}\label{forward}
u_k & = & \argmax\limits_{i\in\mathcal{S},i\notin s_{k-1}} I(\dot{Z}; V(\dot{Z},\Phi_i)|V(\dot{Z},\Phi_{s_{k-1}})), \notag \\
s_k & = & s_{k-1} \cup u_k
\end{eqnarray}
where $k=1,...$, is the iteration index, $u_k$ is the index with the maximum objective function value. Note that $s_0=\emptyset \implies V(\dot{Z},\Phi_{s_0}) =\emptyset$ which reduces the conditional $I(\cdot; \cdot|\cdot)$ to the mutual information $I(\cdot;\cdot)$. During the forward stage, at each iteration $k$ and given the information ($V(\dot{Z},\Phi_{s_{k-1}})$) we already have from the set $s_{k-1}$, we are looking for the function that maximally add extra information to the model. The process terminates when either all basis functions are exhausted (with maximum number of iterations equal $K$), or the reward function $I(\dot{Z}; V(\dot{Z},\Phi_i)|V(\dot{Z},\Phi_{s_{k-1}}))=0$ (or thresholded to $\leq \epsilon$) indicating that none of the remaining basis functions are {\it relevant}, in an information-theoretic sense. In other words, the process terminates when the strongest candidate has no further information beyond what we already have. Note that entropy, mutual information, and conditional mutual information can be estimated using any valid estimator. We choose the $K$-nearest neighbors estimator \cite{kozachenko1987, kraskov2004, vejmelka2008}, for its accuracy, especially with relatively small sample sizes.

After the forward entropic regression, we have the set $s$ that has the indices of the strong candidate functions. Eventually, $s$ may have a few non-relevant functions that are selected due to a high degree of uncertainty and the rounding error at the end of forward entropic regression. Since we have reduced set ($|s|<<K$), it will be computationally inexpensive to validate the accuracy of the model. This backward stage is an elimination stage, where the functions indexed by $s$ are re-examined for their information-theoretic relevance so that redundancy may be removed. In particular, we label the set $s$ as initial set $s_0 = s$ for the backward stage, and we perform the following computations and updates,

\begin{eqnarray}\label{backward}
u_k & = & \argmin_{i\in s_{k-1}} I(\dot{Z}; \mathcal{V}(\dot{Z},\Phi_i)|V(\dot{Z},\Phi_{\{s_{k-1}-i\}})), \notag \\
s_k & = & s_{k-1} / u_k.
\end{eqnarray}
In particular, note the differences of this stage regarding the index set is different between Eq.~(\ref{forward}) and Eq.~(\ref{backward}).
The backward stage includes a loss function, where at each iteration $k$, we examine what information will be lost if we remove the index $i$ from the set $s_{k-1}$  and the process terminates when $I(\dot{Z}; V(\dot{Z},\Phi_i)|V(\dot{Z},\Phi_{\{s_{k-1}-i\}})) > 0$. The result of the backward entropic regression is a set of indices $s$. We emphasize a practical strength, that the forward entropic regression stage can substantially reduce the computational complexity of the backward stage, by limiting the elimination search space to a few candidate functions. However, the backward elimination is key  in that in the special case of a low-dimension system, or we have efficient computational resources, we can plausibly skip the forward stage, to rely only on the backward stage directly, with initial set $s = \{1,2,\dots,K\}$.
Parameters $\mathbf{\beta} \in\mathbb{R}^{K}$, can be found by updating the vector of zeros $\mathbf{\beta}=\mathbf{0}_K$ such that:
\begin{equation}
    \mathbf{\beta}_s = \mathcal{L}(\dot{X},\Phi_s)
\end{equation}
where $\mathbf{\beta}_s$ are the entries of $\mathbf{\beta}$ indexed by the elements of $s$, Thus, $\Vert \mathbf{\beta} \Vert_0 = |s|$. The primary role of entropic regression is to find the minimally optimal informative set of basis functions.  Once that set is identified, the parameter values themselves are easily found, for example by ordinary least squares (OLS).

Computational practice of statistical estimators for information theoretic quantities must be dealt with to make a practical method.
In theory, mutual information $I(x;y|z)$ is always non-negative and $I(x;y|z) = 0$ if and only if $x$ and $y$ are statistically independent given $z$. In practice, however, due to finite sample size and estimation inaccuracies, the estimated mutual information may be nonzero even when $x$ and $y$ are independent, and even worse, some otherwise useful favorite estimators may yield negative numbers \cite{kraskov2004, lord2018}.  We require a way to deduce  with statistical confidence that they differ from zero  to confidently decide whether $x$ and $y$ should be deemed to be independent given the estimated value of $I(x;y|z)$. In \cite{runge2012, sun2015},  a shuffle test  with a ``confidence'' parameter $\alpha\in[0,1]$ for tolerance estimation was developed. The shuffle test involves random shuffling of one of the variables, repeating $n_s$ times, to build a test statistic. In particular, for the $i$-th random shuffle, a random permutation $\pi^{(i)}:[T]\rightarrow[T]$ is generated to shuffle one of the variables, say $y$, which produces a new variable $(\tilde{y}^{(i)})$ where $\tilde{y}^{(i)}=y_{\pi^{(i)}}$; $x$ and $z$ are kept the same. Then, we estimate the mutual information $I(x;\tilde{y}^{(i)}|z)$ using the (partially) permuted variable $(x,\tilde{y}^{(i)},z)$, for each $i=1,\dots,n_s$. For given $\alpha$, we then compute a threshold value $I_\alpha(x;y|z)$ as the $\alpha$-percentile from the values of $I(x;\tilde{y}^{(i)}|z)$. If $I(x;y|z)>I_\alpha(x;y|z)$, we determine $x$ and $y$ as dependent given $z$; otherwise independent. This threshold (tolerance), of mutual independence is adopted in the forward and backward stages as the termination condition.
Hence, the tolerance describes the minimum effective quantity of information. In this sense, in forward entropic regression we are selecting the functions which add a significant quantity of information to the model, while in the backward entropic regression, we discarding functions which are determined to be negligible.

\subsection{Kuromoto Oscillators}
\begin{figure*}[htbp] 
\includegraphics[width=0.95\textwidth]{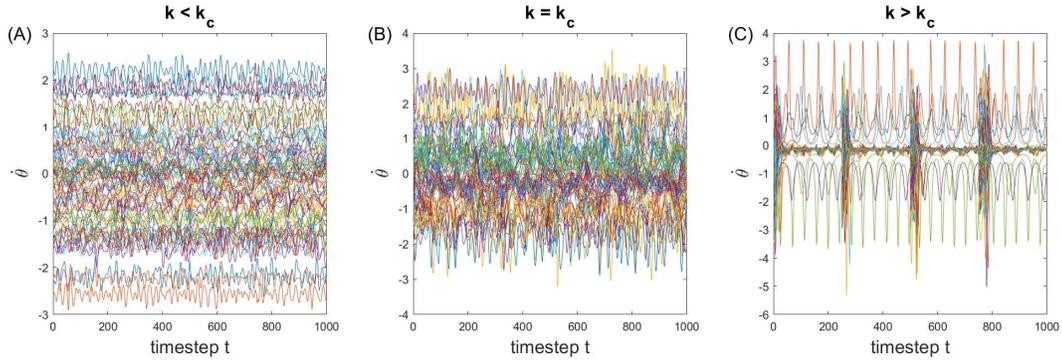}
\caption{The angular frequency $\dot{\theta}$ vs time from a Kuramoto Network model coupled with the DTI Adjacency matrix shown in Fig. \ref{fig:NetworkExamples}-(C). For values of the coupling strength $k_c$ that are below the critical value $k_c$ as in (A) the natural frequencies of the oscillators $\omega$ which are drawn from a normal distribution $\mathcal{N}(0,1)$ have more variance than the mean field of the model. This results in loosely interacting individual oscillators. If the coupling $k$ is greater than the critical value $k_c$ then the oscillators quickly fall into a periodic stable state. (B) Shows the coupling that was chosen for our model. In this mode the oscillators can be shown to be just inside the basin of attraction for the stable synchronous state which is shown in (C), however in practice even without the re-sampling this transient state is long lived when $k$ is chosen to be $k_c$. Note that when $k>k_c$ the oscillators are rapidly attracted to the synchronous basin. \label{fig:Kuramoto}}
\end{figure*}

The Kuromoto model is a  popular model of network coupled phase oscillators \cite{arenas2008}. While in the standard Kuromoto model the network is assumed to be a complete  network \cite{arenas2008}, but this is easily generalized  by \cite{arenas2008,rodrigues2016}:
\begin{equation}
\dot{\theta}_i = \omega_i + k \sum \limits_{j = 1}^n a_{ij}\mbox{sin}(\theta_j - \theta_i), i=1,...n, \label{eq:Kuramoto}
\end{equation}
where $\theta_i(T)$ represents the phase angle of the $i^{\text{th}}$ oscillator  at time $q$, and $\omega_i$ represents the natural frequency and $a_{ij} \in A$. The $\omega_i$ are drawn from some distribution $g(\omega)$ that is usually assumed to be unimodal and symmetric about the mean value $\bar{\omega}$ \cite{arenas2008}.
Let $\boldsymbol{\theta}_i:= \boldsymbol{\theta}(T_i)=(\theta_1(T_i), \theta_2(T_i), ..., \theta_n(T_i)),$ for time $T_i, $ $i=1,..,q$ equally spaced times.
The resulting trajectories data  $Z = (\boldsymbol{\theta}_1,\boldsymbol{\theta}_2,...,\boldsymbol{\theta}_q)^T$, is a $q \times n$, matrix of vector states as columns. The first order Kuramoto model will eventually result in a synchronous state almost certainly \cite{rodrigues2016} if the value of coupling strength $k$ is large enough.  So, if $k>k_{c}$, the result will most likely synchronize for any initial condition. In Fig. 
\ref{fig:Kuramoto}, we show three different scenarios around the critical coupling strength ($k_c$). As $k \rightarrow \infty$, \cite{rodrigues2016}, the system tends to synchronize faster making the system indistinguishable sooner. 

\section{Results}

To demonstrate the effectiveness of entropic regression, we perform analysis on synthetic and semi-synthetic time series data. We know the ground truth for the underlying structural networks so we can compare the effectiveness of entropic regression to other standard leaders based on the correlation method and LASSO regression.
\subsection{Synthetic Data}
\begin{figure*}[htbp] 
\includegraphics[width=0.95\textwidth]{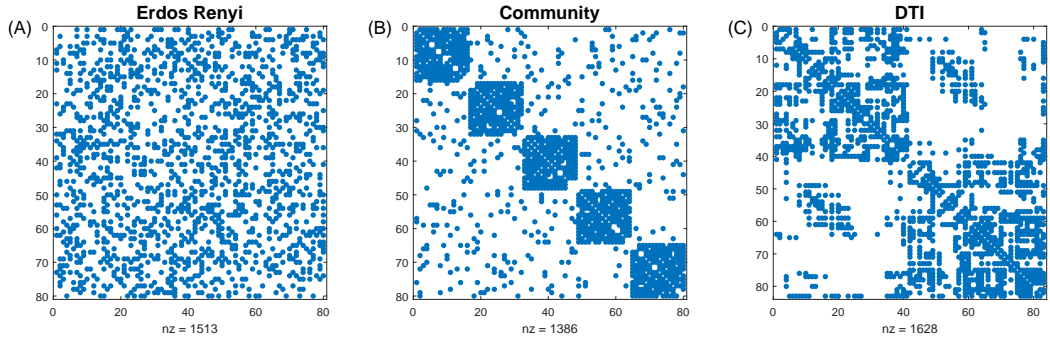}
\caption{We show an example spy plot of the adjacency matrices of the various networks used in this work. The blue dots represent nonzero values of the matrix. As can be seen the ER networks have edges assigned in a random manner, whereas small world networks are highly structured. The DTI network exists somewhere in between these two network types. \label{fig:NetworkExamples}}
\end{figure*}

For the testing purposes we simulate data according to the first order network Kuramoto model as described in Eq. \ref{eq:Kuramoto}. We simulate the Kuramoto model different types of networks with approximately 80 nodes (the DTI network is 83 nodes, the other two types are set to 80 to be close to the same number of nodes), including the directed ER network, a special type of small world network and the DTI network, examples of which are shown in Fig. \ref{fig:NetworkExamples}. The small world network is a simple representation of the brain as seen in \cite{stoltz2017} and the DTI network as a realistic representation. We note that all methods perform poorly if the coupling coeffecient $k$ is set too high as a result of near immediate synchronization, thus rendering the nodes indistinguishable. Thus the coupling strength is chosen to be nearby to its critical value, that is $k \approx k_c$.
\paragraph{}
We ran 50 trials for each of the randomly generated types of adjacency matrix. The parameters of the Kuramoto model where held fixed throughout all trials. The critical coupling of the system $k = k_c$  is defined as, $k_c = \frac{2}{\pi g(0)} \frac{1}{\lambda_{1}}N$ \cite{arenas2008}. The function $g(\omega)$ denotes the pdf of $\omega$ across the network of oscillators, and $\lambda_{1}$ is the largest eigenvalue of the corresponding adjacency matrix $A$, \cite{rodrigues2016}. The initial conditions $\mathbf{\theta_1}$ are randomly drawn from $\mathcal{U}[0,2\pi]$, and the natural frequencies $\omega$ are drawn from $\mathcal{N}(0,1)$. Each sample was integrated by an adaptive stepsize for target precision RK45 numerical solver, and solutions at  1001 equally spaced time steps were computed in the time range $0\leq t\leq 1001$. Finally, $\theta$ is returned as modulated such that $0 \leq \theta < 2\pi$. The derivative $\dot{\theta}_i$ is estimated by finite differences, yielding $1000$ estimated differences. For the Erd\H{o}s-R\'enyi graph \cite{erdos1960}, Fig. \ref{fig:NetworkExamples}(A), the sparsity was chosen such that it was as close as possible to the DTI graph $(~0.2363)$. This, along with choosing a coupling that defined in terms of the adjacency matrix structure are experimental controls available to eliminate  sparsity as a contributing factor in the outcome of the estimation. The Community graph Fig. \ref{fig:NetworkExamples}(B) is created with by adding edges randomly with equal weight as well, however groups are chosen such that each node has a community. We chose 80 nodes with 5 communities of 16 nodes each. Each node has some number of intra-community connections and inter-community connections. We chose 13 intra-community connections and 5 inter-community connections. To generate these graphs efficiently we only guaranty that these are the maximum number of edges and choose the edges with one forward pass. 13 and 5 were chosen so that the community graph has a maximum sparsity of 0.225 as close as possible to the DTI graph. The Community graph is undirected which increases the odds of not finding a configuration with the maximum number of edges, and as such they tend to be sparser than the Erd\H{o}s-R\'enyi graphs.

\begin{figure*}[htbp] 
\includegraphics[width=0.95\textwidth]{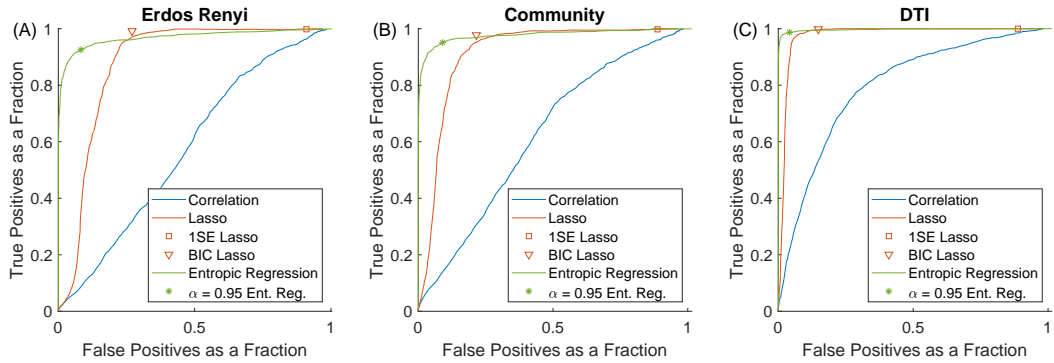}
\caption{Here we show the ROC curves comparing the performance of LASSO, correlation and entropic regression on three different network types. The solid lines represent the ROC curves and the star represents the parameter chose by the entropic regression method, and the triangle represents the parameter $\lambda$ chosen by the BIC method for LASSO.  In all three types of network including the small world network as discussed in \cite{stoltz2017}, it is clear that entropic regression outperforms both correlation and LASSO. Furthermore both entropic regression and LASSO offer significant performance benefits over correlation.  As a result the performance of entropic regression is much better than that of both LASSO and correlation.  \label{fig:ROC1}}
\end{figure*}

\begin{figure*}[htbp] 
\includegraphics[width=0.95\textwidth]{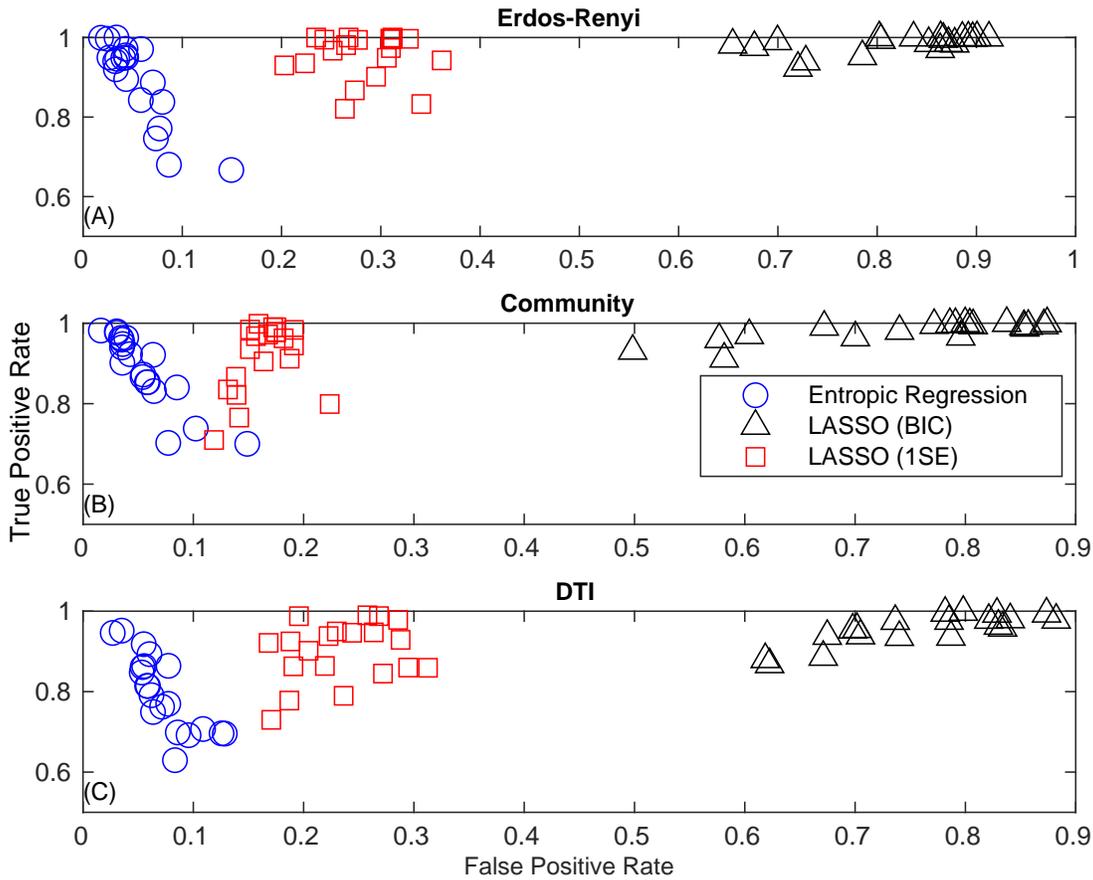}
\caption{Twenty runs, on a different network for each run are shown above. In (A) there are 20 different random Erd\H{o}s-R\'enyi networks with Kuramoto dynamics. The true positive rates and false positive rates are shown for entropic regression and LASSO choosing the appropriate value of $\lambda$ either with BIC or using 1 standard error. (B) Random community networks and (C) DTI networks are used in the place of ER networks from (A). The DTI networks are reconstructed from 20 different patients in this case. It is clear that entropic regression clusters closest to the upper left hand corner, showing its improved performance over the other methods. These results are summarized in \ref{tab:TPR}.  \label{fig:ROC2}}
\end{figure*}

\paragraph*{}
In Fig. \ref{fig:ROC1} we show receiver operator characteristic (ROC) curves for a single realization of the network coupled dynamics discussed above, in which true positive and false positive rates are displayed together. It is desirable in an ROC curve for a method to come as close as possible to the upper left hand corner, representing almost no false positives and almost all true positives. In the three examples, networks from the Erdos Renyi example, the community example show in Fig.~\ref{fig:NetworkExamples}. and the DTI derived example, we compare the accuracy of standard methods such as correlation and LASSO to that of our entropic regression for these three scenarios.  Clearly the entropic regression out performs correlation method.  
In the case of LASSO, the ROC curves are calculated across of a range of values of the regularization parameter $\lambda$. 
We see that for some $\lambda$, that entropic regression clearly outperforms Lasso, but for some $\lambda$ is it close to a tie, with Lasso slightly outperforming entropic regression at least for the Erdos Renyi example.  However, without a method of model selection, or otherwise knowing the answer apriori, it is not possible to know what is the better $\lambda$, the two leading methods being based on either cross-validation or a AIC/BIC formalism already described, and we choose BIC optimum value $\lambda_{opt}$ as discussed above by Eqs.~(\ref{bic1})-(\ref{bic2}).
Assuming errors are Gaussian and i.i.d, therefore the term $\mbox{ln}(\mathcal{L}(\beta,\lambda|X))$ can be approximated as:
\begin{equation}
    \mbox{ln}(\mathcal{L}(\beta,\lambda |X)) = -\frac{q}{2}\mbox{ln}(2q\pi \mbox{MSE}),
\end{equation}
where $\mbox{MSE}$ denotes the mean squared error. We arrive at this approximation by noting that in the Gaussian case we have $\mbox{MSE} = \frac{\sigma^2}{q}$ \cite{degroot1986}. We combine this with the well known Gaussian log-likelihood \cite{zwiernik2014} in the i.i.d. case and ignoring a constant term that does not  effect on the optimization of BIC. This yields:
\begin{equation}
    \mbox{BIC} = \kappa \mbox{ln}(q) + q \mbox{ln}(2q\pi \mbox{MSE}). \label{eq:NewBIC}
\end{equation}
Estimates of the $\mbox{MSE}$ follow a  LASSO solution and using ten-fold cross validation and optimization estimated for  Eq. \ref{eq:NewBIC}. 
\paragraph*{} 
We can see that entropic regression exceeds the performance of the other two methods as the method identifies parameters much closer to the upper left hand corner of the ROC curve than the other methods. For LASSO we show the BIC  estimation of the parameter $\lambda_{opt}$, which is marked by a triangle in Fig. \ref{fig:ROC1}. A common choice of confidence level is $\alpha = 95\%$, which is the point chosen by entropic regression and marked by a star.
It is clear that entropic regression significantly out performs the other two methods. 
\paragraph*{}
{ Fig. \ref{fig:ROC2} shows the TPR and FPR over 20 runs on different networks of the types shown in Fig. \ref{fig:ROC1}. 20 ER and community networks were generated at random, while the DTI networks were reconstructed from 20 different patients. Only the point chosen by $\alpha = 0.95$ for entropic regression and $\lambda$ chosen by BIC or one standard error (1SE) in LASSO are shown, rather than the full ROC. Entropic regression in this case clusters closest to the top left corner in all network types, making it preferable to LASSO. Across the 20 runs, the mean TPRs of LASSO (BIC) and entropic regression are similar, slightly favoring LASSO. Entropic regression significantly outperforms LASSO (BIC) in FPR, similar to above.
\paragraph*{}
As seen in Fig. \ref{fig:NetworkExamples} graphs produced by DTI of real brains lie somewhere between ER and community in terms of ordered structure. For this reason the synthetic fMRI data is more likely to match the dynamics exhibited by the brain. In all 20 examples of simulated dynamics from different DTI networks, entropic regression had a lower false positive rate than any of the networks produced by LASSO. This allows for increased confidence in the edges which are inferred by entropic regression in the context of the brain. This highlights the utility of our approach over other existing methods in neurological applications. \paragraph*{}
LASSO in all cases averaged more than triple the FPR of entropic regression over the 20 networks examined. This could have severe implications for the network tomography. For example a future treatment relying on accurate knowledge of connectivity between the ROIs would suffer greatly from having a significant number of false inferred edges.}

\paragraph*{}
In Table \ref{tab:TPR}, we average across 20 sample simulations, and for each with a new sampled randomly generated network and randomly selected initial condition. {The average TPRs and FPRs as well as the standard deviation of the two best methods are reported.   As can be seen, on average entropic regression and LASSO perform similarly in true positive rate (TPR), however entropic regression significantly outperforms, especially in terms of false positive rate (FPR).} Across the different k Overall, entropic regression is capable of recovering the majority of the true networks while only generating very few false edges, while LASSO also recovers the majority of the true network but introduces many more false edges.

\begin{table}[p]
\begin{tabular}{lllll}
 &ER Network & Small World Network & DTI Network\\
Mean TPR Ent. Reg.  & 0.8938 (0.1035)  & 0.8858 (0.0892) & 0.7983 (0.0940)\\
Mean TPR LASSO (BIC) &   0.9538 (0.0571) & 0.9150 (0.086) & 0.8994 (0.0743)\\
Mean FPR Ent. Reg. & 0.0535 (0.0301)  & 0.0551 (0.0305) & 0.0718 (0.0268)  \\
Mean FPR LASSO (BIC) &  0.2844 (0.0410) &  0.1647 (0.0247) & 0.2350 (0.0440)
\end{tabular}

\caption{True and False Positive rates (TPR and FPR) of entropic regression and LASSO averaged over 20 synthentic network realizations of either Erdos-Renyi or Small World Network types, and 20 DTI networks constructed from different patients. On average both LASSO and entropic regression return similar TPR, however entropic regression returns a much lower FPR than LASSO. In the parenthesis the standard deviation is reported. \label{tab:TPR}}
\end{table}

\section{Conclusions}
In this work we have presented a simple model for brain dynamics using Kuramoto oscillators running on synthetic ER and DTI networks. Having access to accurate network structure is essential to understanding the dynamics of any network coupled system. Unfortunately in the brain the ground truth of the network structure is unknown and thus must be inferred. Common methods for inference of these networks from data include LASSO and correlation. We show that a new method, entropic regression, offers improvement upon both LASSO and correlation in terms of accuracy. Specifically entropic regression offers a similar true positive rate with a much improved false negative rate over the other methods. 

\section*{Data Availability}
The data that support the findings of this study are available from the corresponding author upon reasonable request.

\begin{acknowledgments}
E.B. and A.A, and A. D. were supported by the Army Research Office (N68164-EG) and, J.F. and E. B. were supported by DARPA.
\end{acknowledgments}


\end{document}